\documentstyle[aps,preprint]{revtex}
\begin{document}

\title{Nature of order from random two-body interactions}


\author{S. Dro\.zd\.z and M. W\'ojcik}

\address{Institute of Nuclear Physics, PL--31-342 Krak\'ow, Poland \\    
Institut f\"ur Kernphysik, Forschungszentrum J\"ulich, D-52425 J\"ulich, 
Germany}
\date{\today}
\maketitle

\begin{abstract}
We investigate the origin of order in the low-lying spectra of many-body
systems with random two-body interactions. 
Our study based both on analytical as well as on numerical arguments
shows that except for the most $J$-stretched states, the
ground states in the higher $J$-sectors are more orderly and develop 
larger energy gaps than the ones in the $J=0$-sector. 
Due to different characteristic energy scales in different $J$-sectors
the $J=0$ ground states may predominate only when all the states 
are taken together.
\end{abstract}


\newpage

\section{Introdution}

Nature of the mechanism generating order out of randomness constitutes
one of the most fundamental issues of the contemporary physics.  
Theories based on various versions of ensembles of the random matrices
provide one possible theoretical frame for studying such effects.
The Gaussian orthogonal ensemble (GOE) constitutes the most common
reference.
The related concepts originate~\cite{Wigner,Mehta} from nuclear physics
and prove very useful in the area of strongly interacting Fermi 
systems or in quantum chaos~\cite{Zelev,Guhr}. At present 
they finds even broader applications in such diverse fields
like the brain research~\cite{Kwapien}, 
econophysics~\cite{Laloux,Plerou,Drozdz1} and most recently in the 
"Real-World" networks or graphs~\cite{Farkas}.      
Utility of the standard random matrix theory (RMT) results form the fact that 
a potential agreement reflects the generic properties of a system 
and thus in many cases it provides an appropriate null hypothesis.
From this perspective the deviations are even more interesting 
as they can be used to quantify some properties which are nonrandom 
and thus system specific.     

In this context the recently identified~\cite{Johnson,Bijk} preponderance 
of the $J=0$ ground states in strongly interacting Fermi systems, 
such as atomic nuclei, arising from random two-body interactions seems to 
indicate the effect reflecting a 'sparser connectivity' 
than just pure random.  
Several closely related issues have also been 
addressed in the context of mesoscopic~\cite{Jacquod} and randomly
interacting many-spin systems~\cite{Kaplan}.
One purpose of the present investigation is to identify the origin 
of the related physically relevant deviations from standard RMT
and to quantify their character. Since it was nuclear physics which 
gave birth to RMT we believe that the present example, even though addressed
in the nuclear context, may also stimulate much broader activity 
and understanding of similar effects in other areas.  

\section{Statistics of matrix elements}

Our theoretical framework is thus analogous to this of ref.~\cite{Johnson}. 
Then schematically, indicating nevertheless all the relevant ingredients,
the interaction matrix elements $v^J_{\alpha,\alpha'}$ of good total
angular momentum $J$ in the shell-model 
basis $\vert {\alpha} \rangle$ can be expressed as follows~\cite{Talmi}:
\begin{equation}
v^J_{\alpha,\alpha'} = \sum_{J'} \sum_{i i'} 
c^{J \alpha \alpha'}_{J' i i'}
g^{J'}_{i i'}.
\label{eqv}
\end{equation}   
The summation runs over all combinations of the two-particle states 
$\vert i \rangle$ coupled to the angular momentum $J'$ and connected 
by the two-body interaction $g$. 
$g^{J'}_{i i'}$
denote the radial parts of the corresponding two-body matrix elements while 
$c^{J \alpha \alpha'}_{J' i i'}$ 
globally represent elements 
of the angular momentum recoupling geometry. 
Structures analogous to eq.~(\ref{eqv}) can be identified in various other
areas. The quantum open systems~\cite{DTW}
or the neural nets~\cite{Hopfield} provide immediate examples.

In statistical ensembles of matrices the crucial factor determining the
structure of eigenspectrum is the probability distribution $P_V(v)$ of matrix
elements~\cite{Drozdz2}. Especially relevant are the tails of such distributions
since they prescribe the probability of appearance of the large matrix
elements. From the point of view of the mechanism producing the energy gaps
they are most effective in generating a local reduction of dimensionality
responsible for such effects. In principle, the probability distribution of
the shell model matrix elements is prescribed by their general structure 
expressed by the eq.~(\ref{eqv}), provided the probability distributions
of both $g^{J'}_{i i'}$ and $c^{J \alpha \alpha'}_{J' i i'}$ are known.
In general terms this structure can be considered to be of the form
\begin{equation}
V = \sum_{i=1}^N V_i
\label{eqV}
\end{equation} 
and each $V_i$ to be a product of another two variables denoted 
as $C_i$ and $G_i$. 
By making use of the convolution theorem~\cite{Bracewell} the probability 
distribution $P_V(v)$ 
that $V$ assumes a value equal to $v$ can be expressed as:
\begin{equation}
P_V (v) = 
F^{-1} [F(P_{V_1}(v_1)) \cdot F(P_{V_2}(v_2)) \cdot ... \cdot F(P_{V_N}(v_N))],   
\label{eqPV}
\end{equation}
where $F$ denotes a Fourier transform, $F^{-1}$ its inverse and 
$P_{V_i}(v_i)$ the probability distributions of individual terms. 
Taking in addition into account the fact that 
\begin{equation}
P_{V_i} (v_i) = \int dg_i P_{G_i} (g_i) P_{C_i} ({v_i \over g_i}) 
{1 \over {\vert g_i \vert}}
\label{eqPVi}
\end{equation}
one can explicitely derive the form of $P_V (v)$ in several cases.  
Assuming for instance that all the above constituents are identically
Gaussian distributed (then, according to eq.~(\ref{eqPVi}),
$P_{V_i}(v_i) = K_0 (\vert v_i \vert) /\pi$ and thus
$F(P_{V_i}(v_i)) = 1/\sqrt {1 + \omega^2}$ ) one arrives at
\begin{equation}
P_V (v) = { { \vert v \vert^{(N-1)/2} K_{(N-1)/2} (\vert v \vert) } \over
{ 2^{(N-1)/2} \Gamma(N/2) \sqrt{\pi} }},
\label{eqPr}
\end{equation}
where $K$ stands for the modified Bessel function.
Asymptotically, for large $v$, this leads to 
\begin{equation}
P_V(v) \sim \exp (-\vert v \vert)~{\vert v \vert}^{N/2-1}.
\label{eqPa}
\end{equation}

For such a global estimate the identical Gaussian  distribution of
$g^{J'}_{i i'}$
is consistent both with the Two-Body Random Ensemble (TBRE)~\cite{French}
and with the Random Quasiparticle Ensemble (RQE)~\cite{Johnson}.
The only anticipated difference originates from the fact that in the 
second case the variance of the distribution drops down with $J'$ like
the inverse of $2J' + 1$ which is expected to result in a smaller effective $N$
as compared to TBRE. By contrast, 
in both versions of the above random ensembles the geometry expressed by  
$c^{J \alpha \alpha'}_{J' i i'}$ 
enters explicitely. However, the complicated
quasi-random coupling of individual spins is believed~\cite{Ericson} to
result in the so-called geometric chaoticity~\cite{Zelev}. 
For the extreme values of $J$ the underlying selection rules may however
impose severe constraints in achieving such a limit.
Below we therefore explicitly verify its range of applicability.   
 
\section{The model and results}

The model to be quantitatively explored here consists, 
similarly as in ref.~\cite{Johnson}, of 6 identical particles 
(all single-particle energies are set to zero) operating in
the $sd$ shell.  
Fig.~1 shows distributions of the corresponding
geometrical factors 
$c^{J \alpha \alpha'}_{J' i i'}$ 
for $\alpha \ne \alpha'$
and for several relevant values of $J$ and $J'$.

As one can see, the Gaussian
may be considered quite a reasonable representation of the distribution of such
factors for all combinations of $J$ and $J'$ shown, 
with one exception, for those
which involve $J=0$. In this later case the distribution of    
$c^{0 \alpha \alpha'}_{J' i i'}$ resembles more a uniform distribution 
over a finite interval located symmetrically with respect to zero.
One principal reason for this fact is that the $6j$ symbols which enter
$c^{J \alpha \alpha'}_{J' i i'}$ are here more selective.
These empirical facts justify well the estimates of $P_V(v)$ based on 
eq.~(\ref{eqPr}) for $J \ne 0$ and not so well for $J=0$.   
More appropriate in this particular case is to assume a uniform distribution
of $c^{0 \alpha \alpha'}_{J' i i'}$ over an interval confined by say
$-c_0$ and $c_0$, i.e., $P_{C_i}(c_i) = 1/2c_0$, retaining $P_{G_i}(g_i)$
in its original Gaussian form of course.  
By making use of eqs.~(\ref{eqPV}) and (\ref{eqPVi}) one then obtains
\begin{equation}
P_V(v) = {1 \over \pi} \int_0^{\infty} 
\Bigl [ {\sqrt {\pi \over 2}} {{{\rm erf}(c_0 \omega/\sqrt{2})} \over {c_0 \omega}} \Bigr ]^N
\cos (\omega v) d\omega
\label{eqPVu}
\end{equation}
which for large $v$ behaves like
\begin{equation}
P_V(v) \sim \exp(-{\vert v \vert}^2).
\label{eqPVul}
\end{equation}

An explicit calculation of the distribution of the shell model 
off-diagonal matrix elements for the various $J$-values based on the present
model with two-body matrix elements drawn from RQE (TBRE results in similar
relations among different $J$-sectors though the distributions are somewhat
broader as compared to RQE) confirms the above analytical estimates as is 
illustrated in Fig.~2.

Indeed, such a distribution in the $J=0$ sector resembles more a Gaussian
and the large $v$ tails of this distribution drop down faster as compared 
to the $J \ne 0$-sectors where this asymptotics is exponential
(eq.~(\ref{eqPa})). 
At the same time
the $J \ne 0$ sectors are dominated by very small matrix elements to a larger
degree than $J=0$. The probability of
appearance of a large off-diagonal matrix element which in magnitude
overwhelms the remaining ones is thus greater for $J \ne 0$ than for $J=0$.
Such an effective reduction of the rank in the former case is expected 
to result in a stronger tendency to localization as compared 
to GOE~\cite{Drozdz2,Cizeau}. The corresponding characteristics can be 
quantified in terms of the information entropy
\begin{equation}
K_l^J = - \sum_{\alpha=1}^{M_J} 
\vert a_{l,\alpha}^J \vert^2  \ln \vert a_{l,\alpha}^J \vert^2    
\label{eqE}
\end{equation}
of an eigenstate labelled by $l$ from the $J$-sector. The coefficients
$a_{l,\alpha}^J$ denote the eigenvector components in the basis $\vert\alpha>$. 
Such a mean field basis offers an appropriate reference~\cite{Drozdz3}
for the present purpose.
Since the definition of $K_l^J$ involves the total number of states $M_J$
which differ for different $J$'s, before relating the result to the GOE 
we normalise $K_l^J$ to the GOE limit of this quantity~\cite{Izrail}
\begin{equation}
K_{GOE}^J = \psi (M_J /2 +1) -\psi(3/2),
\label{eqEGOE}
\end{equation}
where $\psi$ is the digamma function.
Within our model the so-calculated and RQE ensemble averaged quantity
for all the states versus their corresponding energies $E_l^J$ is illustrated
in Fig.~3. As anticipated, it is not $J=0$ whose lowest eigenstate comes
out most localised, i.e., most regular. The lowest states for several higher
$J$ values (like 2 and especially 4) deviate much more from GOE. 
This thus indicates more favorable 
conditions for the emergence of energy gaps for larger $J$ than for $J=0$.

Fig.~3 provides one more information which turns out helpful to properly 
interpret the results.
The $J=0$ states are spread over the broadest energy interval even though
the number of states $(M_0=14)$ is here significantly smaller than for 
several larger $J$ values $(M_1=19, M_2=33, M_3=29, M_4=26)$.
As a result, the average level spacing is a factor of few
larger for $J=0$ than for the remaining ones.

In Fig.~4 (dashed line) we therefore show
distributions of the ground state $(E_1^J)$ gaps 
\begin{equation}
s^J=(E_2^J - E_1^J)/D^J, 
\label{eqrho}
\end{equation}
where similarly as in ref.~\cite{Johnson}, 
\begin{equation}
D^J=<E_3^J - E_2^J>, 
\label{eqD}
\end{equation}
though here for each $J$ individually.

As it is clearly seen the $J=0$-sector does not significantly distinguish
from the remaining ones.
In view of our investigations presented above one would
however expect a reduced probability for occurance of the large
ground state energy gaps in this particular sector. 
As the solid lines in Fig.~4 indicate such an effect
does indeed take place when $D^J$ in eq.~(\ref{eqrho}) is replaced by
\begin{equation}
{\bar D}^J = <E_{M_J}^J - E_2^J>/(M_J-2).
\label{eqDbar}
\end{equation}  
In fact, it seems more appropriate and more consistent with the above global
considerations to relate the ground state energy gap just to the average
global level spacing among the remaining states, 
characteristic for a given $J$, as expressed by eq.~(\ref{eqDbar}).

Finally one may ask a question why this tendency does not extend to the
highest $J$-values. In this connection one has to remember that the
off-diagonal matrix elements is not the only relevant element.
These are the diagonal matrix elements which constitute the driving term.
Irrespective of the value of $J$ their distribution is always Gaussian-like.
This can be observed numerically and is consistent with arguments formulated
in terms of eqs.~(\ref{eqv} - \ref{eqPa}) since the geometrical factors 
$c^{J \alpha \alpha'}_{J' i i'}$ entering the diagonal matrix elements are
always nonnegative. As it is shown in Fig.~5 (dashed lines), 
increasing however $J$ beyond 4 results in a significant 
reduction of the variance of $P_V(v)$ for the diagonal matrix elements 
and consequently a larger fraction of the off-diagonal matrix elements becomes 
effective in mixing the basis states. In addition, due to a smaller number 
of terms entering the eq.~(\ref{eqv}), for the stretched high-$J$
states the effect of geometric chaoticity is no longer effective and
in this respect the conditions become similar to those for $J=0$.
As a result, the distribution of off-diagonal matrix elements
converts back towards more Gaussian-like shaped, i.e. $N_{eff}$ becomes larger
(for $J=6$ not shown in Fig.~2 $N_{eff}=2.67$). Superposing the above
two effects one thus obtains even smaller gaps and 
even more delocalized states at the edges of the spectra as compared to $J=0$.
In fact the spectral density (solid lines in Fig.~(5)) becomes even somewhat 
closer to semicircular in this case.

\section{Summary}

The present investigation based both on theoretical as well
as on numerical arguments clearly shows that the many-body problems 
described in terms of various variants of the two-body random ensembles
(like RQE or TBRE) develop quantitatively well identified deviations from
the GOE. These deviations can be linked to differences in the distribution 
of matrix elements and quantified in terms of the localisation or of
energy gaps in eigenspectra. Contrary to the common belief
they point to the intermediate total angular momenta
as those $J$-sectors whose ground states are ordered most.
From this perspective, a predominance of the $J=0$ ground states~\cite{Johnson}
can be viewed as a result of mixing states with different 
characteristic energy scales from different $J$-sectors.  
It seems also appropriate to notice here that the
arguments formulated in terms of eqs.~(\ref{eqv}-\ref{eqPa}) provide a more
adequate approach towards understanding the distribution of matrix elements
in realistic nuclear shell-model calculations than the ones based on 
multipole expansion~\cite{Zelev}. Finally, similar parallels between 
the distribution of matrix elements and the structure of eigenspectra 
relative to the GOE can be set in the recent econophysics~\cite{Drozdz4} 
applications.  

We acknowledge useful discussions with J.~Oko{\l}owicz, M.~P{\l}oszajczak
and I.~Rotter at the early stage of this development. This work was 
partly supported by KBN Grant No. 2 P03B 097 16 and by the German-Polish
DLR scientific exchange program, grant No. POL-028-98.

\newpage

\begin{center}
{\bf FIGURE CAPTIONS}
\end{center}
{\bf Fig.~1} The normalised distribution of geometrical factors  
$c^{J \alpha \alpha'}_{J' i i'}$ entering the off-diagonal matrix elements
(eq.~(\ref{eqv})) for the model of 6 particles in the $sd$-shell. \\ 
{\bf Fig.~2} The probability distributions of nonzero many-body off-diagonal
matrix elements in different $J$-sectors drawn from one thousand of RQE
samples of two-body matrix elements. The energy scale is set by $\bar v$,
where $w_{J'} = {\bar v}^2/(2J'+1)$ and $w_{J'}$ determines the RQE 
mean square variance. These distributions are fitted (solid lines)
in terms of eq.~(\ref{eqPr} with $N$ treated as a fitting parameter.
The corresponding best $N$'s $(N_{eff})$ for each $J$ are listed.
By increasing $N$ the distribution prescribed by eq.~(\ref{eqPr}) quickly
approaches (as a consequence of the central limit theorem) the Gaussian
distribution. In this way the $J=0$ distribution is demonstrated to be much 
closer to the Gaussian than the remaining ones whose asymptotic behaviour 
is consistent with a slower, exponential fall-off. \\
{\bf Fig.~3} The information entropy normalised to its GOE limit
$(K_l^J / K_{GOE}^J)$ for all the states $l$ from various $J$-sectors
(all positive parity)
versus energies $(E_l^J)$ of those states. All the quantities are ensemble
averaged. The energy units are the same as in Fig.~2. \\   
{\bf Fig.~4} Distributions of ground state energy gaps $s^J$ as 
defined by the eq.~(\ref{eqrho}) for successive $J$'s.
The dashed line uses $D^J$ defined by 
eq.~(\ref{eqD}) while the solid line the one defined by eq.~(\ref{eqDbar}). \\
{\bf Fig.~5} The probability distributions of the diagonal
matrix elements in different $J$-sectors drawn from one thousand of RQE 
samples of two-body matrix elements (dashed lines) and the corresponding 
spectral densities (solid lines). The energy units are the same as in Fig.~2.
 
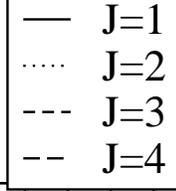
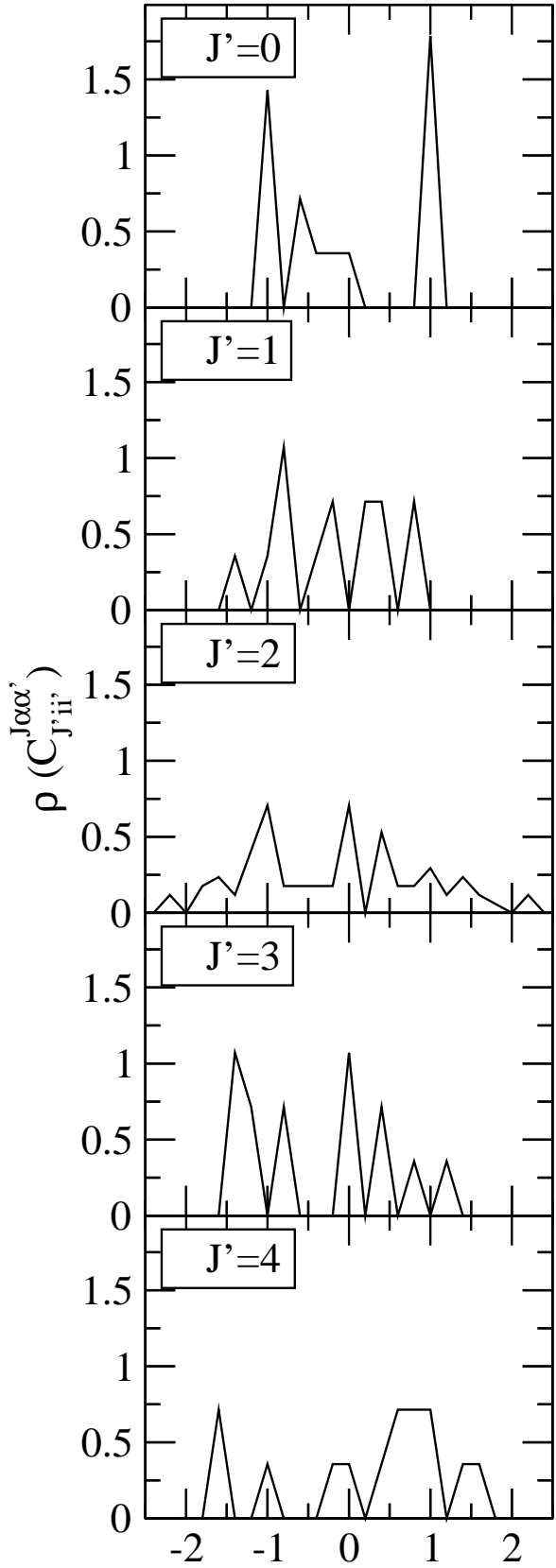
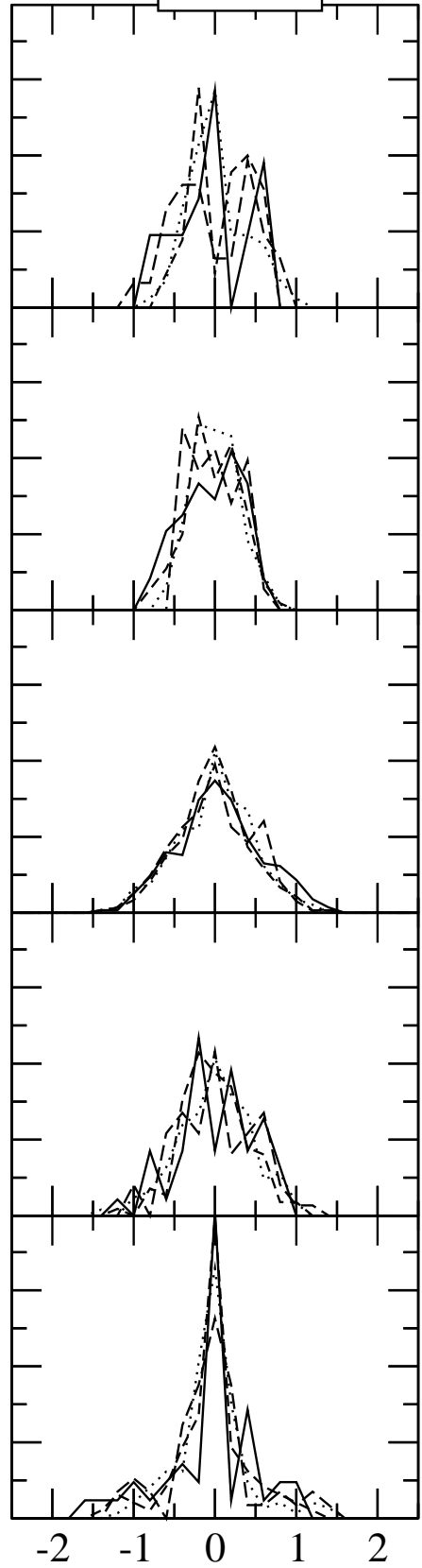
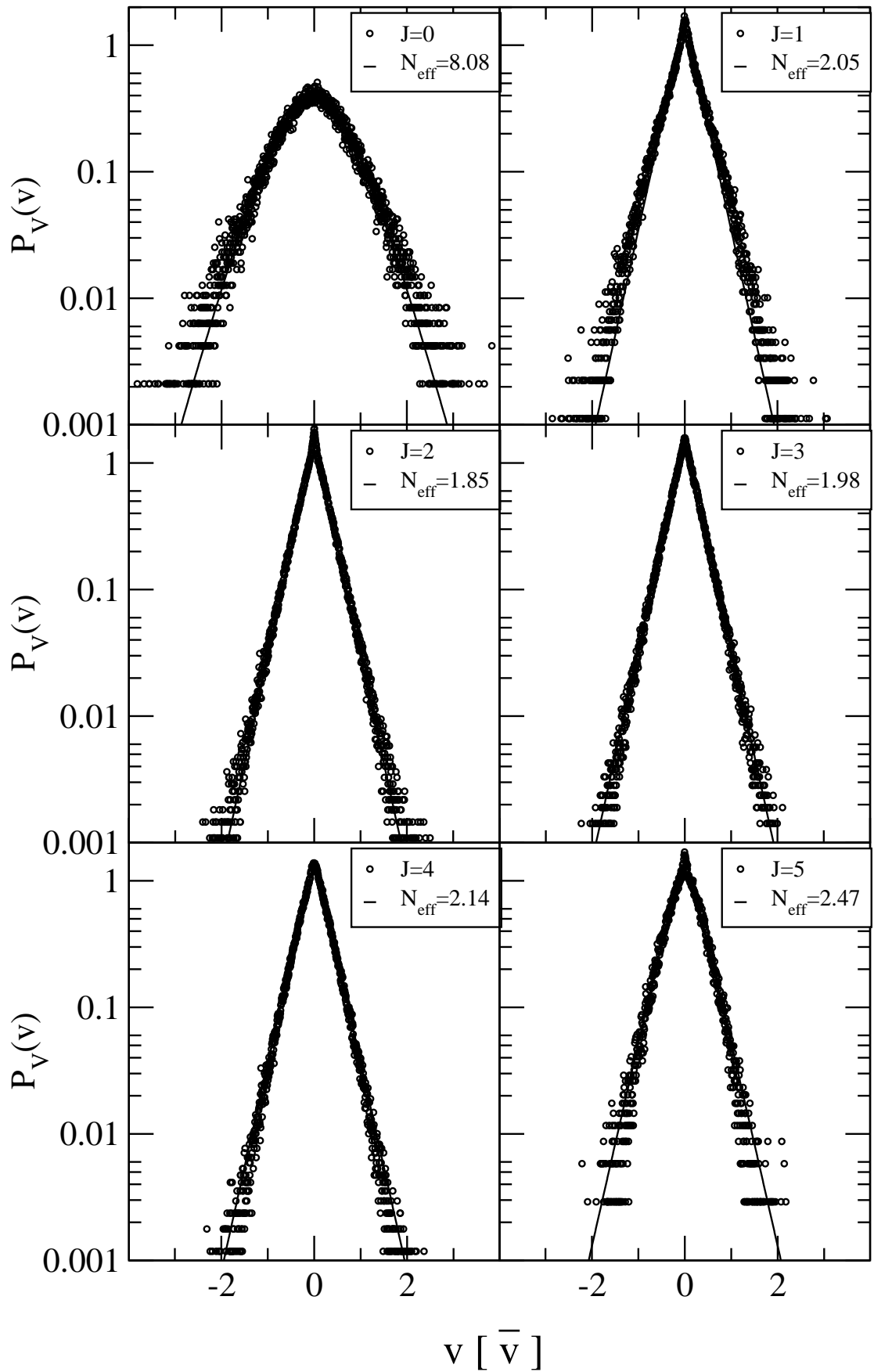
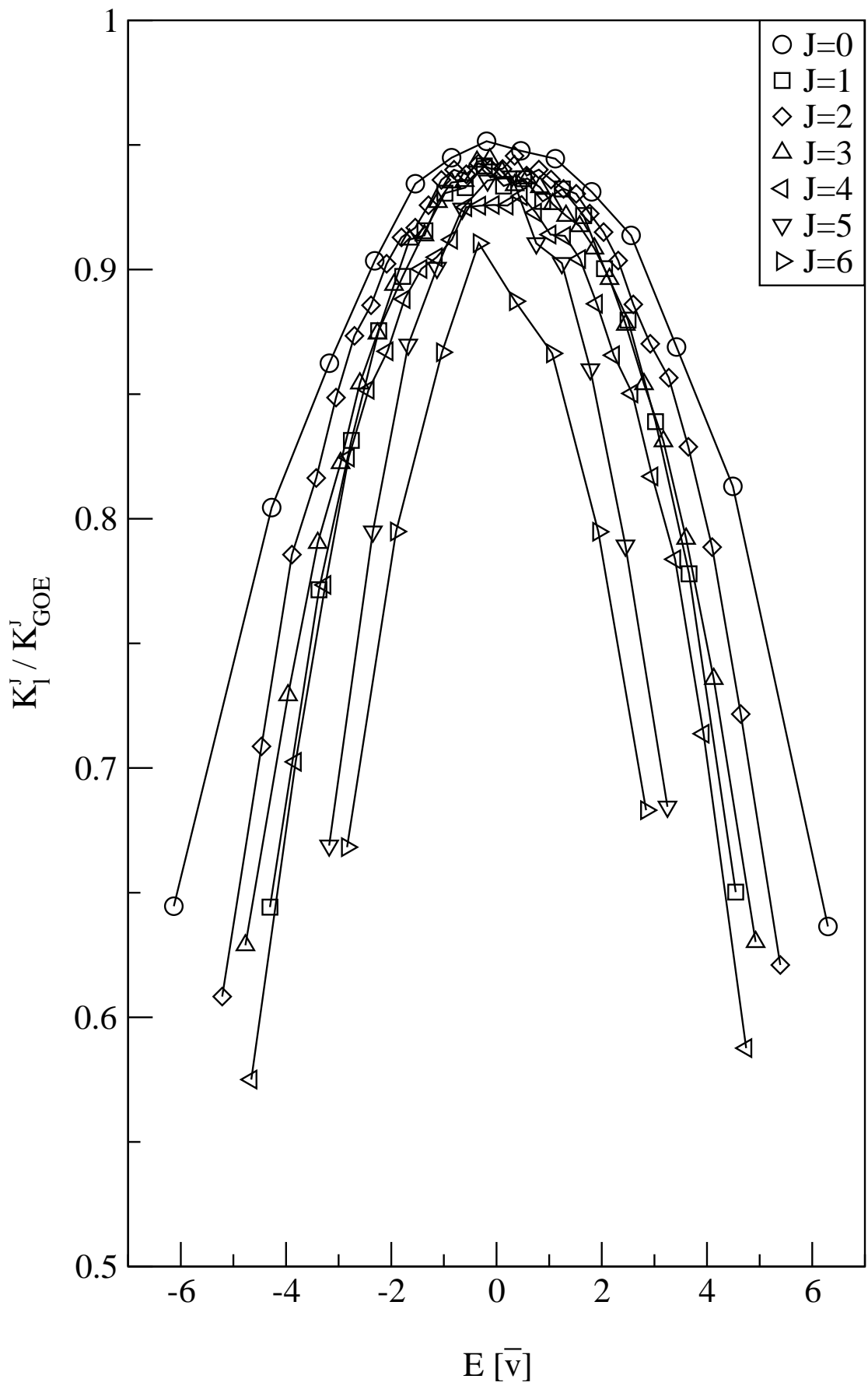
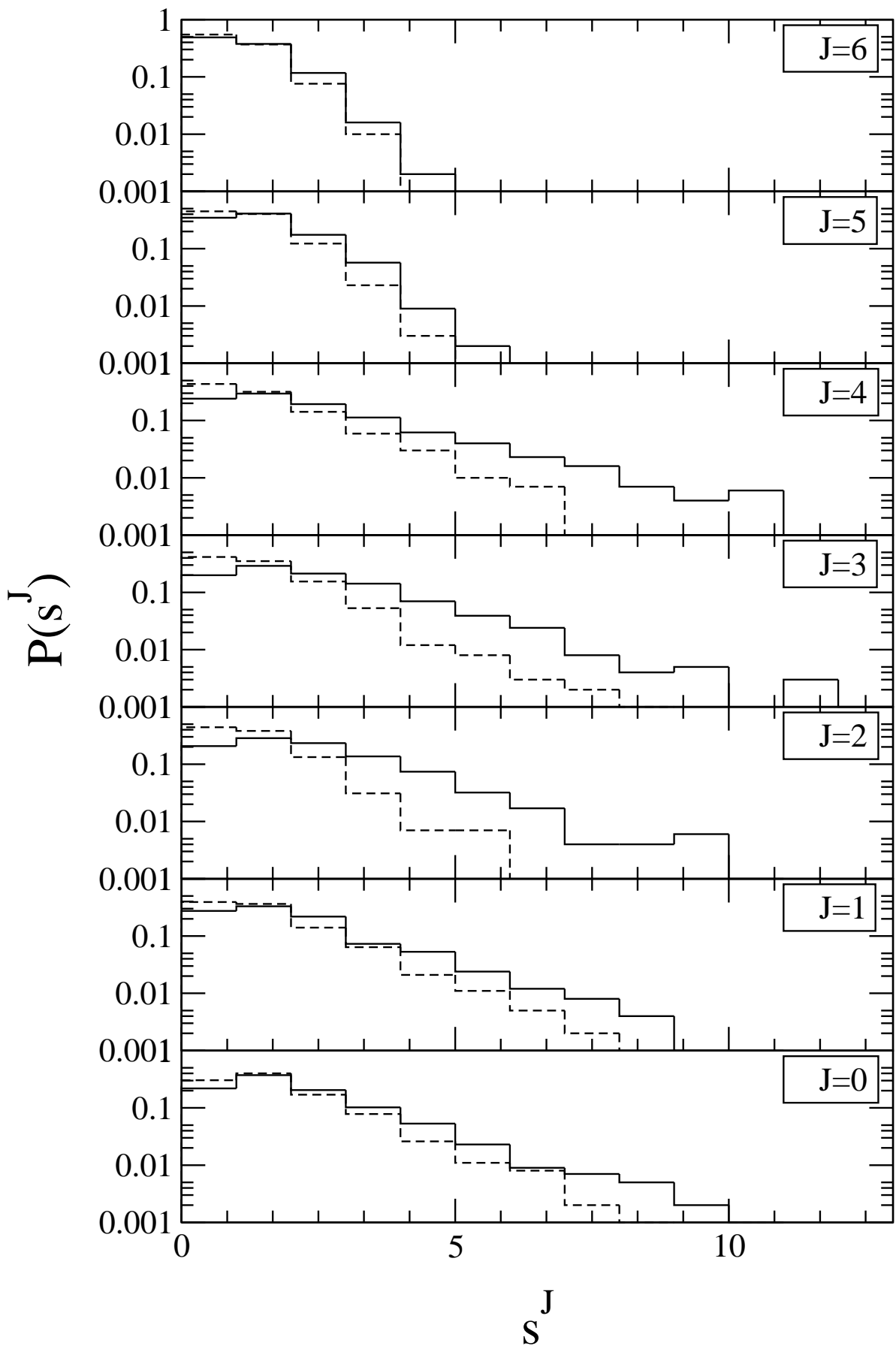
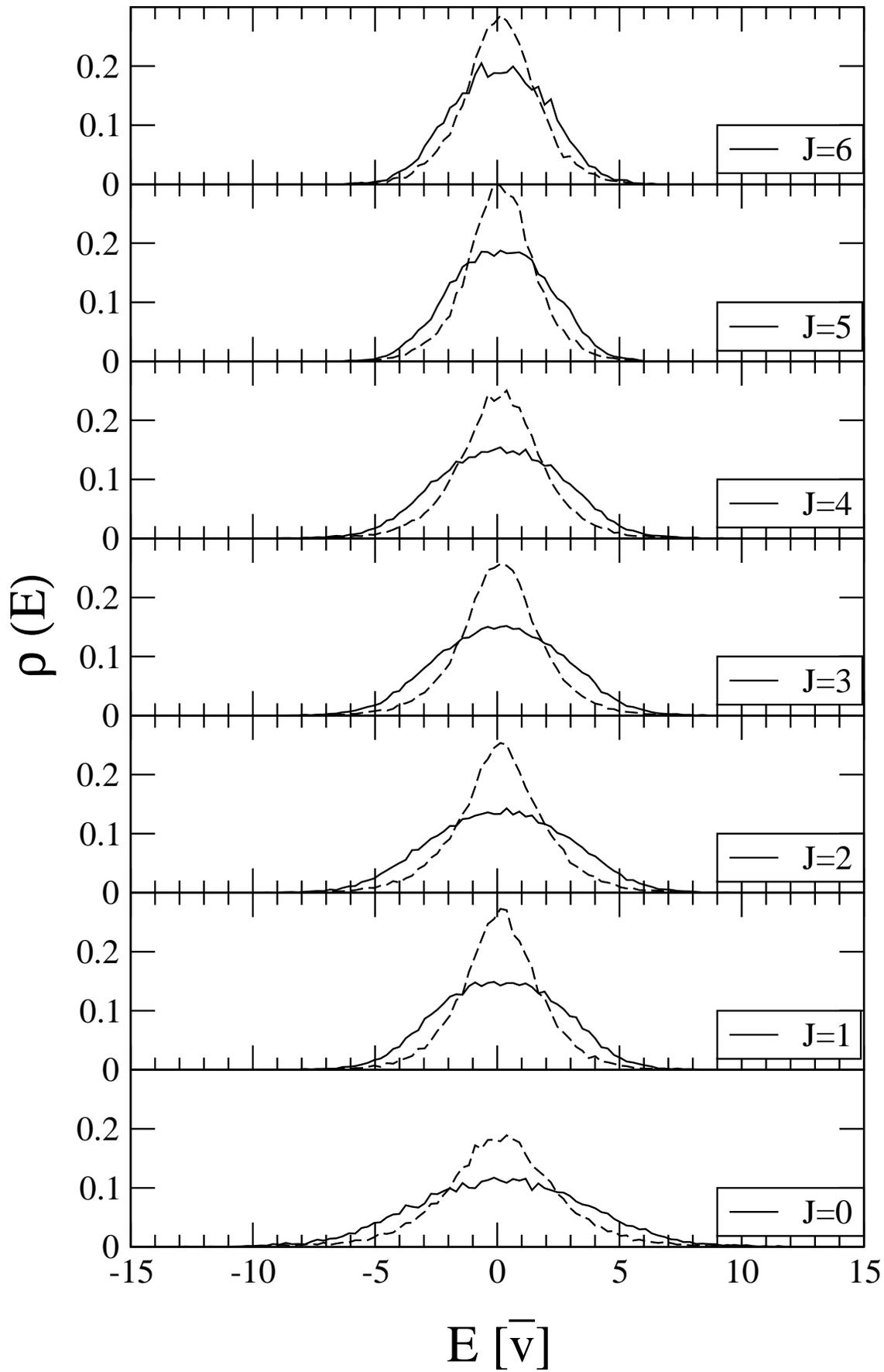
\end{document}